\documentclass[%
 reprint,
superscriptaddress,
 amsmath,amssymb,
 aps,
]{revtex4-2}

\usepackage{amsmath}
\usepackage{multirow}
\usepackage{array}
\usepackage{graphicx}
\usepackage{bm}
\usepackage{hyperref}
\usepackage{orcidlink}
\usepackage{cleveref}
\usepackage{xspace}
\usepackage{booktabs}
\usepackage[commentmarkup=uwave]{changes} 

\usepackage{siunitx}

\usepackage{bm}
\def\mathbf#1{\bm{#1}}

\setdeletedmarkup{\color{red}\sout{#1}} 
\definechangesauthor[color=red   , name={Matt Duez}         ]{MD}
\definechangesauthor[color=green , name={Courtney Cadenhead}]{CC}
\definechangesauthor[color=blue  , name={Zach Etienne}      ]{ZEreword}
\definechangesauthor[color=brown  , name={Z Etienne}     ]{ZE}
\definechangesauthor[color=orange, name={Bernard Kelly}     ]{BK}
\definechangesauthor[color=purple, name={Leo Werneck}       ]{LW}

\def\etal{\textit{et al}.\xspace}
\def\form#1#2{\ensuremath{{}^{#1}\!#2}}
\def\Eqref#1{Equation~(\ref{#1})\xspace}
\def\Eqrefp#1{Eq.~\ref{#1}\xspace}

\newcommand{\WSU}{\affiliation{Department of Physics \& Astronomy, Washington State University, Pullman, Washington 99164, USA}}

\newcommand{\UIdaho}{\affiliation{Department of Physics, University of Idaho, Moscow, ID 83844, USA}}

\newcommand{\Maryland}{\affiliation{Center for Space Sciences and Technology, University of Maryland Baltimore County, 1000 Hilltop Circle Baltimore, MD 21250, USA}}

\newcommand{\GoddardA}{\affiliation{Gravitational Astrophysics Lab, NASA Goddard Space Flight Center, Greenbelt, MD 20771, USA}}

\newcommand{\GoddardB}{\affiliation{Center for Research and Exploration in Space Science and Technology, NASA Goddard Space Flight Center, Greenbelt, MD 20771, USA}}

\begin{document}

\title{Toward 2D Dynamo Models Calibrated by Global 3D Relativistic Accretion Disk Simulations}

\author{Matthew D. Duez}\WSU
\author{Courtney L. Cadenhead}\WSU
\author{Zachariah B. Etienne}\UIdaho
\author{Bernard Kelly}\Maryland \GoddardA \GoddardB
\author{Leonardo~R.~Werneck}\UIdaho

\date{\today}

\begin{abstract}
Two-dimensional models assuming axisymmetry are an economical way to explore the long-term evolution of black hole accretion disks, but they are only realistic if the feedback of the nonaxisymmetric turbulence on the mean momentum and magnetic fields is incorporated.  Dynamo terms added to the 2D induction equation should be calibrated to 3D MHD simulations.  For generality, the dynamo tensors should be calibrated as functions of local variables rather than explicit functions of spatial coordinates in a particular basis.  In this paper, we study the feedback of non-axisymmetric features on the 2D mean fields using a global 3D, relativistic, Cartesian simulation from the IllinoisGRMHD code.  We introduce new methods for estimating overall dynamo alpha and turbulent diffusivity effects as well as measures of the dominance of non-axisymmetric components of energies and fluxes within the disk interior.  We attempt closure models of the dynamo EMF using least squares fitting, considering both models where coefficient tensors are functions of space and more global, covariant models.  None of these models are judged satisfactory, but we are able to draw conclusions on what sorts of generalizations are and are not promising.

\end{abstract}

\maketitle

\section{Introduction}
\label{sec:intro}

Untilted black hole accretion disks are three-dimensional, turbulent systems, but their fluid and field profiles can be viewed as combinations of an axisymmetric, slowly evolving ``background'' and non-axisymmetric ``fluctuations.'' Often, only the background is of interest, with fluctuations considered primarily for their feedback effects on the background. This perspective motivates the use of 2D (axisymmetric) simulations to model accretion flows.

Simply evolving the general relativistic magnetohydrodynamic (GRMHD) equations in 2D will almost certainly be inadequate; MHD flows are often not only quantitatively but even qualitatively different in 2D versus\ 3D, as illustrated by opposite turbulent energy cascades~\cite{Batchelor:1969abc,Davidson:2015abc} and the 2D anti-dynamo theorem~\cite{Cowling:1933abc,Moffatt:1978abc}.  What we aim for are 2D evolutions that ``look like'' 3D evolutions.  Note that even this is ambiguous.  Do we want 2D evolutions that resemble the azimuthal average of 3D evolutions, or do we want 2D evolutions that resemble representative 2D slices of 3D evolutions?  Both approaches might be interesting, but they will likely be very different.  Consider an evolution variable $\Psi(r,\theta,\phi,t)$ with azimuthal Fourier decomposition $\sum_m\hat{\Psi}_m(r,\theta,t)e^{im\phi}$.  Azimuthal averaging picks out the $m=0$ contribution, and if this is subdominant, the azimuthal average $\overline{\Psi}$ will be significantly smaller than a 2D slice of a 3D realization.  We might also expect most of the $(r,\theta)$ fluctuation to average out, so that azimuthal averaging might act as a low-pass filter, removing most eddies.  Given these differences, one must choose when designing a model whether the 2D fields represent azimuthal averages or a representative slice.  For this paper, we consider 2D models of azimuthally averaged 3D evolutions.

One might also worry that the distinction of background versus fluctuating turbulence is too sharp.  In terms of the Fourier decomposition of fields in the azimuthal direction, it seems safer to distinguish high-$m$ modes from the background than, for example, $m=1$ or $m=2$ modes.  A possible criterion would be to consider an $m\ne 0$ mode is ``turbulence'' only if the eddy turnover time is less than the orbital period.  For simplicity, consider unmagnetized incompressible turbulence with turbulent velocity ${\sim}v_T$ at the largest length scale $H$ (the disk scale height) and the usual Kolmogorov energy cascade. At scale $\ell$, the velocity scale is $v_{\ell}$ and the timescale is $\tau_{\ell}\approx \ell/v_{\ell}$.  As specific kinetic energy density flows to lower scales at the same rate ${\approx} v_{\ell}^2/\tau_{\ell}$ for all $\ell$, it follows that $\tau_{\ell}\approx H^{1/3}\ell^{2/3}v_T^{-1}$.  Turbulent modes then have $\tau_{\ell}<\Omega^{-1}$, where $\Omega$ is the orbital frequency.  Estimating $m\approx R/\ell$ at radius $R$, the eddy time is then sufficiently short if
\begin{equation}
m > m_{\rm crit}\approx R H^{1/2}(\Omega/v_T)^{3/2}\ .
\end{equation}
It might be necessary to evolve $m<m_{\rm crit}$ (a compromise between 2D and 3D, assuming $m_{\rm crit}$ turns out to be a small number), with the $m>m_{\rm crit}$ contribution safely modeled by a turbulent/dynamo closure.

Filtering away the fluctuating component alters the evolution equations, as can be seen by azimuthally averaging the Newtonian induction equation:
\begin{equation}
  \begin{split}
    \overline{\partial_t\mathbf{B}} &= \overline{\mathbf{\nabla}\times (\mathbf{B}\times\mathbf{v}}) \\
    &= \mathbf{\nabla}\times (\overline{\mathbf{B}\times\mathbf{v}}) \\
    &\equiv \mathbf{\nabla}\times (\overline{\mathbf{B}}\times \overline{\mathbf{v}} + \mathbf{\Delta E})\;.
  \end{split}
\end{equation}
To evolve these equations, we require a closure condition, providing the dynamo EMF $\mathbf{\Delta E}$ as a function of azimuthally averaged variables.  A commonly considered closure~\cite{Moffatt:1978abc,Brandenburg:2004jv} is
\begin{equation}
  \Delta E^i = \alpha^i_j B^j + \eta^i_j J^j \label{eq:NewtonianDynamo}
\end{equation}
where $J^j$ is the current, $B^j$ is the magnetic field, $\alpha^i_j$ is the dynamo alpha tensor, and $\eta^i_j$ is the turbulent magnetic diffusivity tensor.

Dynamo corrections to 2D GRMHD simulations have been undertaken by a number of groups~\cite{Sadowski:2014awa,Bugli:2014via,Tomei:2019zpj,Vourellis:2021yru,Shibata:2021xmo}, usually assuming an isotropic dynamo:  $\alpha^i_j=\alpha_{\rm dyn}\delta^i_j$, $\eta^i_j=\eta_{\rm dyn}\delta^i_j$.  The $\alpha_{\rm dyn}$ term enables the dynamo alpha effect:  toroidal field produces poloidal field.  The alpha effect enables 2D GRMHD simulations to maintain magnetic fields at the strength seen in 3D simulations.  (Note that using it for this reason implies a representative 2D slice 
rather than an azimuthal average interpretation.)  The $\eta_{\rm dyn}$ term
is the turbulent magnetic diffusivity, which acts like a resistivity.  As $\alpha_{\rm dyn}$ is a pseudoscalar, it is expected to switch sign at the equator, so some such spatial dependence on $\alpha_{\rm dyn}$ is often stipulated. 

Dynamo terms are also sometimes added to 3D GRMHD simulations~\cite{Giacomazzo:2014qba,Vigano:2019,Carrasco:2019uzl,Most:2023sme}.  In this case, the distinction between background and fluctuation (and hence the meaning of averaging) is done differently---the fluctuation is the MHD flow at spatial scales below the grid scale.  It is not clear \textit{a priori} if the same $\mathbf{\Delta E}$ model should apply to the two cases.

Three-dimensional MHD simulations can be used to extract $\alpha^i_j$ and $\eta^i_j$.  Gressel and Pessah~\cite{Gressel:2015mxa} calculated these using a test-field method. They studied shearing box simulations and averaged over boxes, so their distinction of mean versus fluctuating fields was not quite the same as the azimuthal average, but they were able to study dependence on shear and vertical field. Assuming $\alpha^i_j=\alpha_{\rm dyn}\delta^i_j$ and $\eta^i_j=0$, Hogg and Reynolds~\cite{Hogg:2018zon} used global MHD simulations to extract $\alpha_{\rm dyn}$ in each hemisphere. Bendre~\etal~\cite{bendre2020turbulent} introduced a method of extracting $\alpha^i_j$ and $\eta^i_j$ by least-squares fitting via singular value decomposition (SVD), which they applied to radiatively inefficient accretion disks in Dhang~\etal~\cite{Dhang:2019kqo}. Both test-field and SVD methods find deviations in dynamo tensors from isotropy and report fairly consistent magnitudes. Dhang~\etal note the difficulty of extracting a clear dynamo signal in the disk region, as well as other indications that the field and flow are mostly non-axisymmetric.  The SVD method has also been used for solar dynamo and binary neutron star remnant calculations~\cite{racine2011mode,Kiuchi:2023obe}.

The recent study by Jacquemin-Ide~\etal~\cite{JacqueminIde:2023qrj}, using GRMHD disk simulations to study dynamo action from the nonlinear evolution of the magnetorotational instability, also presents some relevant findings.  They reported that $m=1$ to $m=3$ modes are crucial in dynamo generation of poloidal field (perhaps suggesting a suitable $m_{\rm crit}$). 
They also emphasize the importance of magnetic flux advection from the outer to the inner disk.  This might be a less prominent effect in the small disks (resembling short gamma ray burst setups) often studied in compact binary merger contexts.  Finally, they do estimate the viability of an $\alpha_{\rm dyn}$ and $\eta_{\rm dyn}$ dynamo closure using correlation coefficients between $\Delta E_{\phi}$ and $B_{\phi}$ or $J_r$.  The average $\Delta E_{\phi}$-$B_{\phi}$ correlation oscillates about zero and does not show a significant time-average; the $\Delta E_{\phi}$-$J_r$ correlation does have a time average clearly distinct from zero, but is strongest with a time offset between cause and effect, suggesting it is mediated by additional dynamics.

The number of studies that extract dynamo coefficients from global disk simulations, particularly relativistic simulations, remains small given the parameter space of disk states, and not all existing results have high statistical significance.  Furthermore, dynamo coefficient extraction is of two types:  either a single average coefficient is extracted for the entire disk (or a hemisphere thereof), or dynamo coefficients are extracted at each $(r,\theta)$ point in space.  These spatial dependencies are presumably proxies for dependence on the local physical quantities and their derivatives.  If these dependencies could be made explicit, the resulting model for $\Delta E^i$ with only a few global fitting constants (as opposed to fitting scalar functions of space) could be expressed in covariant tensor form, making it immediately generalizable to general spatial coordinate systems and more easily generalizable to other accretion systems (varying disk thickness, magnetic flux, and rotation law).

In this paper, we present a new analysis of the azimuthally averaged evolution of a magnetorotationally turbulent disk around a Kerr black hole.  We use data from the Cartesian numerical relativity code IllinoisGRMHD~\cite{Etienne:2015cea,Noble:2005gf}.  We present a detailed analysis of the degree of non-axisymmetry, introducing measures of its dominance in magnetic energy, kinetic energy, and angular momentum transport.  We find that most of the magnetic field and non-azimuthal velocity field are averaged out by azimuthal averaging, and momentum transport is predominantly non-axisymmetric. Thus, azimuthally averaged 3D ideal GRMHD resembles viscous hydrodynamics more than 2D ideal GRMHD.  We also propose ways of estimating an average $\alpha_{\rm dyn}$ and $\eta_{\rm dyn}$.  An elegant definition of the former comes from the Lorentz invariants of the azimuthally averaged field tensor.  For the latter, we take advantage of the fact that, at least for the first \num{10000}\,$M$, 2D ideal GRMHD overpredicts the magnetic energy compared to azimuthally averaged 3D GRMHD, so we estimate $\eta_{\rm dyn}$ by the resistivity needed to achieve this level of field suppression.  For this purpose, we add a new phenomenological resistivity to the HARM code and compare with the 3D results.

Next, we attempt to extract $\alpha^i_j$ and $\eta^i_j$ using SVD least-squares fitting.  We consider models for which coefficients are functions of $(r,\theta)$, functions of $\theta$, or functions of local scalars and pseudoscalars (with spacetime dependency only from those scalars and pseudoscalars).  We introduce norms for the error in the best-fit model and variance in coefficient extraction to easily assess the reliability of models.  None of the models meet all of our pre-set standards.

Since these models assuming \Eqref{eq:NewtonianDynamo} are not judged satisfactory, we assess certain alternative classes of models which maintain the same basic structure of the dynamo EMF. 
 First, we consider whether a successful global model could be created if the RMS deviation from axisymmetry of the velocity or magnetic field were known.  To be usable as a closure condition, an evolution equation for one of these variables that tracks the true evolution well enough would have to be introduced, similar to the evolution of the mean turbulent kinetic energy in $k-\epsilon$ models~\cite{Jones:1972abc,Pope:2000abc,Davidson:2015abc} (which might be generalized to MHD~\cite{Kenjerevs:2000abc,Meng:2018ak}).  In fact, such scalars do not provide acceptable dynamo models, so the motivation to devise such evolution equations does not arise. Second, we ask whether the residual EMF subtracting off the EMF computed from $m=0$ to $m=3$ components of $\mathbf{v}$ and $\mathbf{B}$ is more amenable to dynamo models of the form \Eqref{eq:NewtonianDynamo}.  It is not.  We conclude that more general models for $\Delta E^i$ are likely required.

This paper is organized as follows.  In Sec.~\ref{sec:relativistic_dynamo}, we review the GRMHD and dynamo equations.  In Sec.~\ref{sec:nonaxi-mag}, we provide details on the 3D simulation, in particular analyzing the contributions of non-axisymmetric fields.  In Sec.~\ref{sec:isotropic}, we describe how to produce estimates of the isotropic $\alpha_{\rm dyn}$, $\eta_{\rm dyn}$.  In Sec.~\ref{sec:nonisotropic}, we describe how we produce fits to $\alpha^{ij}$, $\eta^{ij}$ and present quality of fits.  We summarize and present conclusions in Sec.~\ref{sec:conclusions}.

We use units $G=c=1$ throughout.  Latin letters from the beginning of the alphabet ($a$--$f$) are spacetime indices, running 0--3.  Indices $i$, $j$, $k$ are spatial, running 1--3.  When writing equations in geometric form, vectors and forms are written in boldface, with 2-forms getting a ``2'' superscript prefix (e.g., \form{2}{\mathbf{B}}).

\section{An analytic dynamo model}
\label{sec:relativistic_dynamo}

For use in a relativistic code, the dynamo closure condition must at least be spatially covariant; this will account for the unpredictable evolution of the coordinate system or the deliberate use of curvilinear coordinates or coordinates with designed radial or angular concentrations.  Azimuthal averaging introduces a preferred direction, the azimuthal Killing vector $\mathbf{e_{\phi}}\equiv \partial/\partial\phi$, which generally will leave an imprint in the $\mathbf{\alpha}$ and $\mathbf{\eta}$ tensors, so that 3D covariant equations for these tensors will be expected to include $\mathbf{e_{\phi}}$.  Whether the equation must be 4D covariant, i.e., Lorentz covariant, is more debatable.  It could be argued that the averaging procedure itself, which is defined to be azimuthal average at a fixed time, breaks Lorentz invariance.  Indeed, subgrid and effective viscosity prescriptions that are purely spatial, which have been widely and successfully used in numerical relativity~\cite{Radice:2017zta}, can have similar justifications.  On the other hand, in one case, we found it important to keep terms first-order in velocity to properly recover the Newtonian limit~\cite{Duez:2020lgq}. 

Assume a foliation of the spacetime with the usual normal to the slice $\mathbf{\tilde{n}} = -N \mathbf{\widetilde{dt}}$ and 3-metric $\gamma_{ab} = g_{ab} + n_a n_b$.  The field tensor associated with the azimuthally averaged fields is
\begin{align}
    F^{ab} &= n^a \overline{E}^b - n^b \overline{E}^a + \epsilon^{abc}\overline{B}_c \;,\\
    \star F^{ab} &= \overline{B}^a n^b - \overline{B}^b n^a + \epsilon^{abc}\overline{E}_c\;,
\end{align}
where $\overline{E}\cdot n = \overline{B}\cdot n = 0$, and $\epsilon^{abc}=n_d\epsilon^{abcd}$.  (Here $\epsilon^{abcd}=|g|^{-1/2}[abcd]$ is the Levi-Civita tensor and $[abcd]$ is the totally antisymmetric Levi-Civita symbol.)  For the rest of this section, we will suppress lines above averaged quantities, assuming that all quantities are azimuthal averages. 

The fluid frame is defined by the azimuthally averaged 4-velocity $\overline{u}^a$, which can be decomposed as \mbox{$u^a \equiv W n^a + \mathcal{V}^a$},
where $W$ is the Lorentz factor and $\mathbf{\mathcal{V}}$, defined so $\mathcal{V}^an_a=0$, is the Eulerian velocity, not the transport velocity.  For the rest of this section, we will suppress lines above averaged quantities, assuming that all quantities are azimuthal averages.

The electric field $\form{u}{E^a}$ in the fluid frame,
which is also the Lorentz force, is
\begin{equation}
   \form{u}{E^a} \equiv F^{ab} u_b = W E^a  + n^a E^b \mathcal{V}_b + \epsilon^{abc}\mathcal{V}_b B_c\;. \label{eq:E_fluid}
\end{equation}
In ideal MHD, this is zero, and the components parallel and perpendicular to $\mathbf{n}$ must independently vanish, so
\begin{equation}
    \form{\rm MHD}{E^a} = -\frac{1}{W}\epsilon^{abc}\mathcal{V}_bB_c\;. \label{eq:E_ideal}
\end{equation}
One can also compute the electric field using the transport velocity 3-vector $v^i\equiv u^i/u^t$ (cf.~\cite{Baumgarte:2002vv}):
\begin{equation}
  E_i = -\frac{1}{N}\epsilon_{ijk}(v^j+\beta^j)B^k\;, \label{eq:E_transport}
\end{equation}
where $N$ is the lapse and $\beta^i$ is the shift.

In the fluid frame, the magnetic field field is
\begin{equation}
\label{eq:uB}
    \form{u}{B^a} = -{\star}F^{ab} u_b
    = W B^a + B^b\mathcal{V}_b n^a - \epsilon^{abc}\mathcal{V}_b E_c\;.
\end{equation}
A covariant expression that recovers the Newtonian limit (Eq.~\ref{eq:NewtonianDynamo}) is
\begin{equation}
\label{eq:dynamo}
    F^{ab}u_b = -\alpha^a_b\, \form{u}{B^b}
    + \eta^a_b \epsilon^{cbde}u_c\nabla_d\, \form{u}{B_e}\;.
\end{equation}
Note that, in the last term, we can switch freely between $\nabla$ and $\partial$ since $d$ and $e$ are antisymmetrized.

Another possibility would have been to use the actual current
\begin{equation}
    F^{ab}u_b = -\alpha^a_b\, \form{u}{B^b}
    + \eta^a_b\, \form{u}{J^b}\;,
\end{equation}
which is different from the previous equation because of the displacement current $\mathcal{L}_{\mathbf{n}}\mathbf{E}$ term in Ampere's law.  However, the identification of the diffusivity term with $\mathbf{\nabla}\times \mathbf{B}$ should be considered more fundamental than its identity with $\mathbf{J}$, because it is motivated by an expansion in terms of one over the length scale of the magnetic field.  Hereafter, we will simply denote the curl of $\mathbf{B}$ as $\mathbf{J}$.

Unfortunately, \Eqref{eq:uB} itself has the electric field in it.  Luckily, this term is proportional to velocity and subdominant in most regions. For a dynamo equation with only an isotropic $\mathbf{\alpha}$ tensor ($\eta^a_b=0$), Most has shown that one can solve for $E^a$ analytically without needing to assume small velocity~\cite{Most:2023sme}.  Instead, we will substitute the ideal MHD electric field \eqref{eq:E_ideal} in the $\mathbf{\eta}$ term, so that the weak dynamo case is properly recovered.  Then
\begin{equation}
\label{eq:uBidealE}
  \begin{split}
   \form{u}{B^a} &= W B^a + B^b \mathcal{V}_b n^a - \epsilon^{abc}\mathcal{V}_b (-W^{-1})\epsilon_{cef} \mathcal{V}^eB^f \\
   &= B^b\mathcal{V}_b n^a + \frac{1}{W}\left[B^a + B^b\mathcal{V}_b \mathcal{V}^a\right]\;,
  \end{split}
\end{equation}
where we have used $W^2=1+\mathcal{V}^2$.
As for the curl term in \eqref{eq:dynamo}
\begin{equation}
    \epsilon^{abcd}u_a\partial_c \form{u}{B_d} = -W \epsilon^{bcd}\partial_c \form{u}{B_d} + \epsilon^{abcd} \mathcal{V}_a \partial_c \form{u}{B_d}\;.
\end{equation}
The first is just a spatial curl, albeit of the boosted $B$, which is easy to compute.  The second term is, in general, difficult to calculate because it will include time derivatives which are less easily available to an evolution code or post-processing analysis.  However, in most cases, this term will be small.  Suppose we are willing to limit ourselves to cases in which $\mathcal{V}^i$ and $n^i$ are small enough (implying that the shift vector is small, which is usually the case when not very close to a black hole for commonly chosen gauge conditions) that we can keep only terms at lowest order in these.  Then $\mathcal{V}_0=B_0=0$, and for spatial components of the free index $b$ the index $c$ must be the one that is the time component.  In the weak-dynamo limit, the time derivative of $B^a$ is proportional to $\mathcal{V}^i$, and for slowly-evolving metrics, $\partial_tB_a$ is proportional to $\partial_t B^b$.  So this term, which looks first-order in $\mathcal{V}$, is actually second-order, and will be dropped with slightly easier conscience.  We also drop the term $B^b\mathcal{V}_b n^a$ in $\form{u}{B^a}$ (see Eq.~\ref{eq:uBidealE}) as second-order.

For the component of Eq.~\eqref{eq:dynamo} perpendicular to $\mathbf{n}$, and using \eqref{eq:E_fluid}, we are left with
\begin{equation}
  \begin{split}
    W E^a = &-\epsilon^{abc} \mathcal{V}_b B_c
    - \alpha^a_b W^{-1} [B^b + B^c \mathcal{V}_c \mathcal{V}^b] \\
     & - \eta^a_b W \epsilon^{bed}\partial_e \left[(B_d + B^c\mathcal{V}_c \mathcal{V}_d)W^{-1} \right]\;.
  \end{split}
  \label{eq:relativistic_dynamo}
\end{equation}

\section{Non-axisymmetric features of the 3D simulation}
\label{sec:nonaxi-mag}

\subsection{Methods of 3D simulation}

The simulation we used was produced as part of the Event Horizon GRMHD code comparison project~\cite{EventHorizonTelescope:2019pcy}.  The fixed spacetime is a Kerr black hole with dimensionless spin parameter $a/M = 0.9375$.  The gas is an ideal gas with adiabatic index $\Gamma=4/3$.  The initial state of the gas has constant specific angular momentum, defined as $\ell = u^tu_{\phi}$.  In the absence of magnetic forces, the gas forms an equilibrium torus with an inner radius of $6M$ and maximum density at $12M$.  We seed a magnetic field at $t=0$, with the maximum of the magnetic pressure $P_B^{\rm max}$ and the maximum of the gas pressure $P_{\rm gas}^{\rm max}$ related as $P_B^{\rm max}/P_{\rm gas}^{\rm max}=100$.  The field is calculated from a toroidal vector potential $A_{\phi}\propto {\rm max}(\rho/\rho_{\rm max}-0.2,0)$.  This initial state produces a ``Standard and Normal Evolution'' (``SANE'') accretion flow, i.e., one for which magnetic flux on the horizon does not significantly affect accretion. 

The IllinoisGRMHD~\cite{Etienne:2015cea,Noble:2005gf} simulation uses a Cartesian FMR grid (utilizing the Cactus/Carpet infrastructure), in which the coarsest grid is a cube with a half-sidelength of $1750M$. There are 7 different resolutions used, including 6 levels of refinement atop this coarse grid, and the finest-grid cube has a resolution of $\Delta x=\Delta y=\Delta z \approx M/4.388571$. The finest grid is composed of four cubes, each with a half-sidelength of $27.34375M$, and is larger than the black hole, which has a radius of $1.348M$. IllinoisGRMHD evolves a vector potential at cell edges to exactly preserve a finite difference version of the constraint $\mathbf{\nabla\cdot B}=0$~\cite{Etienne:2010ui}. The magnetized fluid is evolved using a high-resolution shock-capturing scheme, with PPM reconstruction and an HLLE approximate Riemann solver.

The disk was evolved for a time of $\num{10000}M$, long enough to see many orbits of fully-developed turbulence.  Volume data of all evolution variables were stored on a \mbox{$(N_r, N_\theta, N_\phi) = (150, 75, 64)$} spherical-polar grid.  Data was output at a high cadence of 0.57$M$, but we find it sufficient to use only every $\Delta t=100M$ for time averages below.  When performing fits for dynamo coefficients, we use timesteps separated by $\Delta t=300M$, where the increased spacing---roughly corresponding to the orbital period at the initial location of maximum density---is chosen to reduce correlation between data at adjacent steps while retaining sufficient data (31 steps), starting at $t=\num{1000}M$, after the magnetic field has become strong and the disk transition from the kinematic to the dynamical regime.

The evolution time is admittedly shorter than usual for dynamo studies.  Ordinarily, one would want to observe multiple cycles of the expected dynamo wave.  Assuming an $\alpha \Omega$ dynamo with $\alpha_{\rm dyn}\sim 10^{-4}$, a characteristic vertical scale $H\sim 10M$, and a characteristic orbital frequency $\Omega\sim 10^{-2}M^{-1}$, we would expect a dynamo period ${\sim}\sqrt{H/(\Omega\alpha_{\rm dyn})}\sim 10^4M$~\cite{Brandenburg:2004jv}, meaning the full evolution time is not more than a cycle, and we will not be able to produce ``butterfly'' diagrams of the average toroidal field at different latitudes showing multiple dynamo cycles, as has been done in some longer disk simulations (e.g.~\cite{Hogg:2018zon,Liska:2018btr,Hayashi:2021oxy}). 
We will, however, evolve comfortably long enough to observe $\mathbf{\Delta E}$ acting.

\subsection{Overview of 3D data}

\begin{figure*}
\includegraphics[width=2\columnwidth]{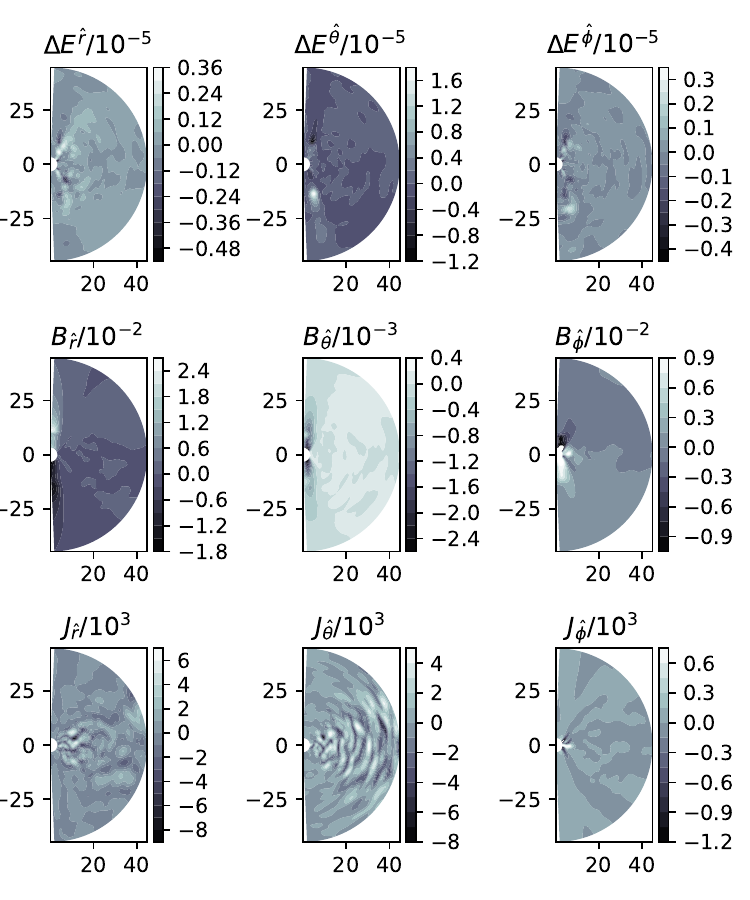}
\caption{Azimuthal and time averages of $\Delta E^{\hat{i}}$, $B_{\hat{i}}$, and $J_{\hat{i}}$. } 
\label{fig:EBJ}
\end{figure*}

We define the following averages
\begin{gather}
\overline{X}(r,\theta,t) \equiv \frac{1}{2\pi} \int_0^{2\pi}  d\phi X(r,\theta,\phi,t)\;, \\
\langle X\rangle(r) \equiv \frac{\int_T^{2T}dt\int_{\pi/3}^{2\pi/3}  d\theta \int_0^{2\pi}d\phi \sqrt{\gamma} X(r,\theta,\phi,t)}{\int_T^{2T}dt\int_{\pi/3}^{2\pi/3} d\theta \int_0^{2\pi}d\phi \sqrt{\gamma}}\;,
\end{gather}
where $T=\num{5000}M$.

To minimize effects of the metric on the spatial dependencies of vector and tensor components, we will use the normed components $X^{\hat{i}}\equiv \sqrt{\gamma_{ii}}X^i$ and $X_{\hat{i}}\equiv \sqrt{\gamma^{ii}}X_i$ when generating plots and fitting formulas. (Note that we drop the Einstein summation convention in the above definitions, so there is no sum over $i$.)

For a quantity $X(x_A)$ which is a nonlinear function of the primitive variables $x_A=(B^i,v^i,\rho,P)$, define the residual as
\begin{equation}
\Delta X = \overline{X} - X(\overline{x}_A)\;.
\end{equation}
Suppose instead of retaining only the average of the spherical-polar components of $\mathbf{v}$ and $\mathbf{B}$, one were to retain the azimuthal Fourier decomposition, truncated at a particular $m=m_t$.  Call $\Psi_{m_t}$ the approximation of $\Psi$ with all azimuthal Fourier modes of $m<m_t$.  Then define the residual EMF $\Delta_m\mathbf{E} \equiv \overline{\mathbf{B}\times\mathbf{v}} - \overline{\mathbf{B}_m\times\mathbf{v}_m}$.  Thus, $\Delta \mathbf{E} = \Delta_0\mathbf{E}$.

\begin{figure*}[!htb]
\includegraphics[width=2\columnwidth]{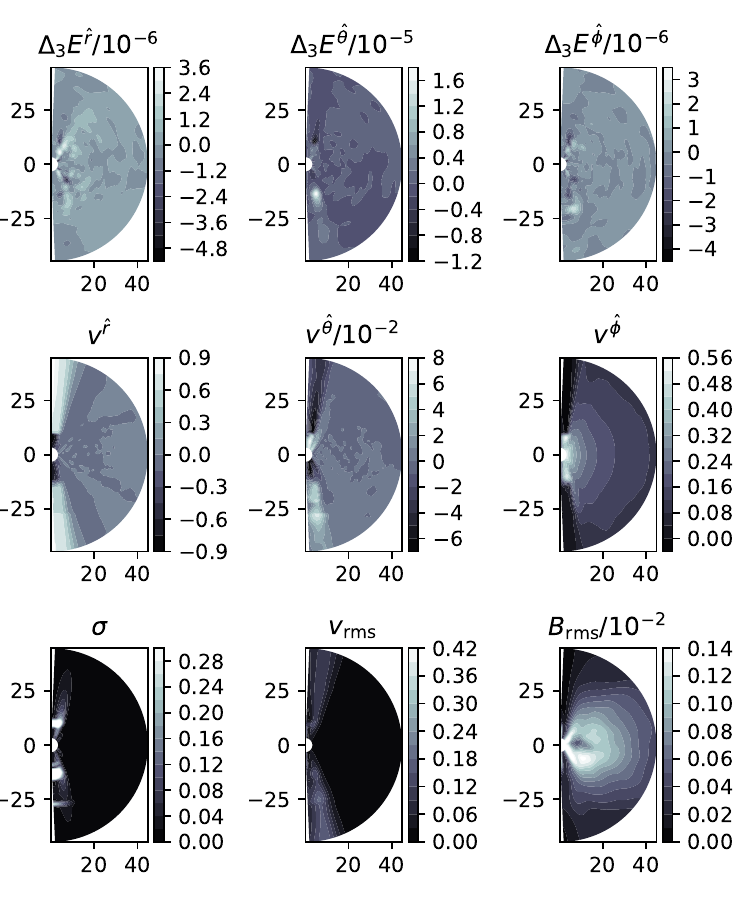}
\caption{Azimuthal and time averages of $\Delta_3E^{\hat{i}}$, $v^{\hat{i}}$, and three scalars:  the 2D shear $\sigma$ and the RMS deviations of velocity and magnetic field from their azimuthal mean (see \Eqrefp{eq:rms}).}
\label{fig:Evcor}
\end{figure*}

\begin{figure}[!htb]
\includegraphics[width=1\columnwidth]{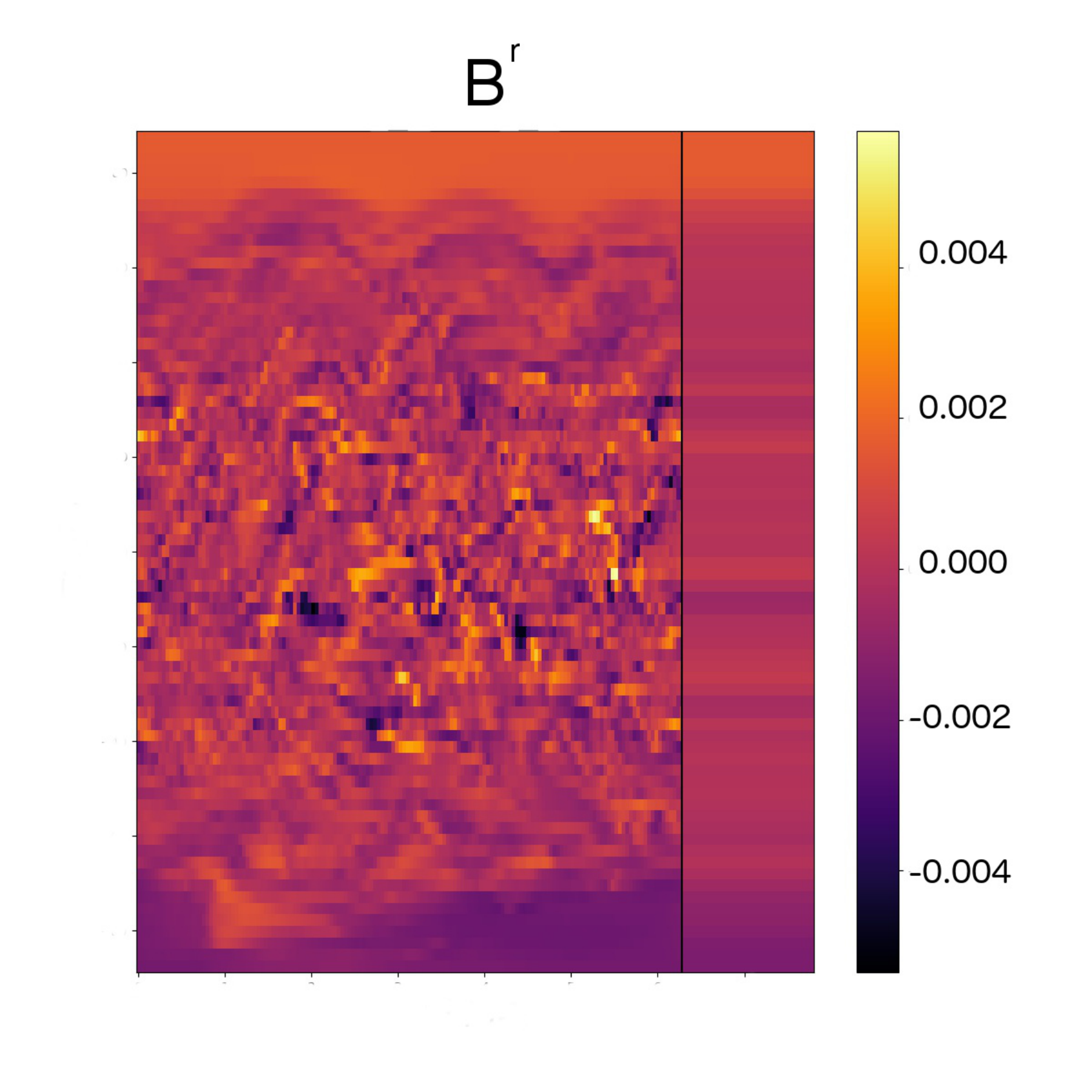} \\
\includegraphics[width=1\columnwidth]{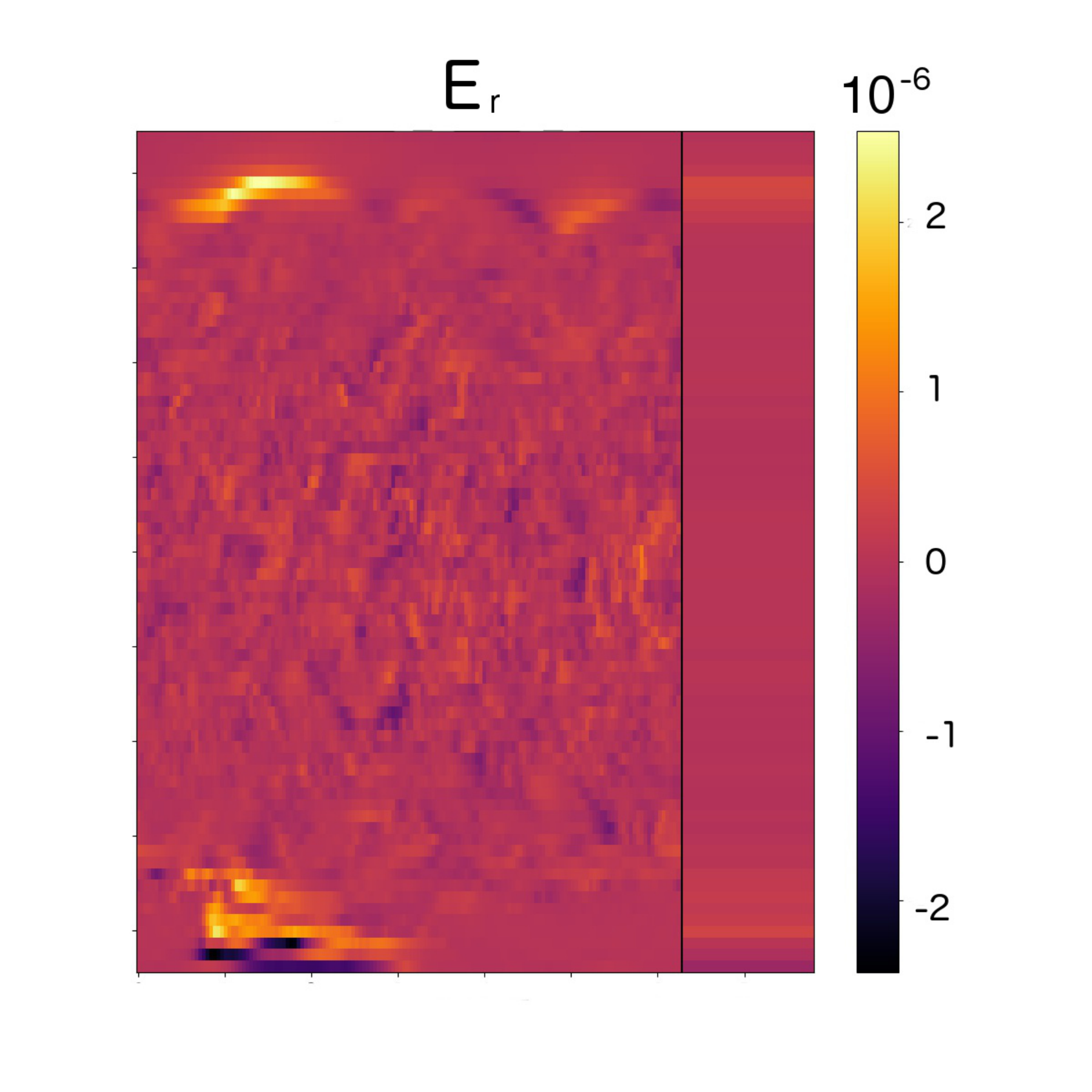}
\caption{Latitude-longitude plots of $B^r$ and $E_r$ at radius $r=25M$ at time $t=\num{5200}\,M$.  As on a standard global map, vertical direction is latitude (with the top corresponding to the North pole and the bottom to the South pole), and horizontal direction is longitude. On the right bar, we show the azimuthal average. }
\label{fig:latlong}
\end{figure}

In Fig.~\ref{fig:EBJ}, we plot azimuthal and time averages of $\Delta E^{\hat{i}}$, $B_{\hat{i}}$, and $J_{\hat{i}}$.  A linear relationship between $\Delta E^{\hat{i}}$ and $B_{\hat{i}}$ or $J_{\hat{i}}$ is not immediately suggested to the eye.  Of the three vectors, the $B_{\hat{i}}$ components are spatially smoothest, while $J_{\hat{i}}$ show small-scale behavior.  $B_{\hat{i}}$ are strongest near the poles, $J_{\hat{i}}$ in the disk, and $\Delta E^{\hat{i}}$ in between.

In Fig.~\ref{fig:Evcor}, we plot azimuthal and time averages of the components of the transport velocity $v^{\hat{i}}$ and the EMF residual beyond $m=3$, denoted $\Delta_3E^{\hat{i}}$ in accord with the notation introduced above.  The structure of $\Delta_3E^{\hat{i}}$ is similar to $\Delta E^{\hat{i}}$, suggesting that the former (representing higher-$m$ modes) significantly contributes to the latter.  Velocity is dominated by the disk rotation and the polar outflow; little of the disk turbulence survives the averaging process.  We also plot three scalar quantities which might be expected to correlate with $\Delta_3E^{\hat{i}}$ (but do not), whose discussion we defer to Sec.~\ref{sec:global_models}.

\subsection{Non-axisymmetric energies and fluxes}

The azimuthal structure and the effect of azimuthal averaging are illustrated in Fig.~\ref{fig:latlong}, in which $B^r$ and $E_r$ are plotted as functions of $\theta$ and $\phi$ at one radius and time.  The azimuthal average is shown as a bar on the right.  We see that, at least for these quantities, azimuthal mean structure dominates only near the poles.  At intermediate latitudes (i.e. in and near the disk), there is small-scale structure (high $m$ contribution), which averages out to value much smaller than the maximum magnitudes. 

\begin{figure}
\includegraphics[width=1\columnwidth]{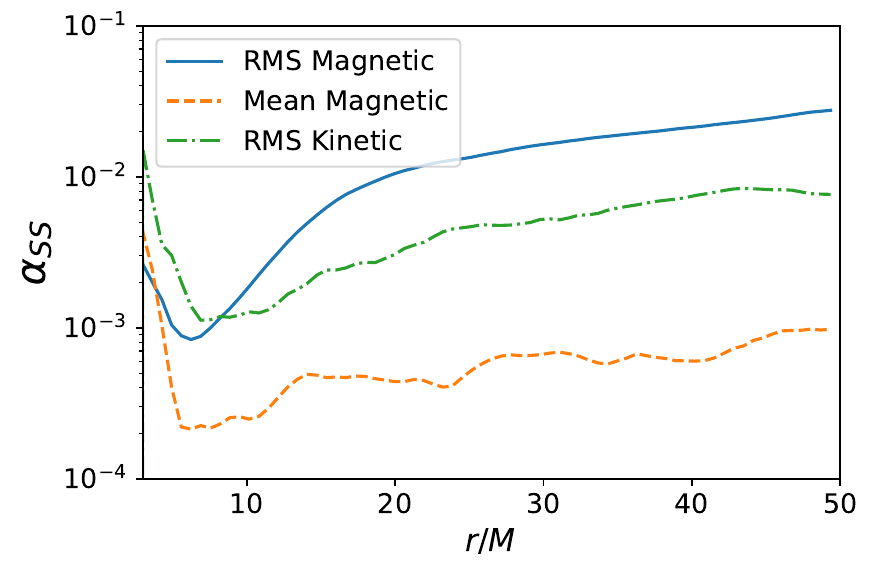}
\caption{Time and angle-averaged momentum fluxes within the disk as functions of $r$.  Magnetic $\alpha_{\rm SS}$ is defined as in Eq.~\ref{eq:alphass_from_avg} using either the mean magnetic field (``Mean'') or the full field with mean subtracted (``RMS''), while RMS kinetic energy flux uses $t^{r\phi}$ instead of $w^{r\phi}$.}
\label{fig:alpha_ss}
\end{figure}

As in~\cite{EventHorizonTelescope:2019pcy}, we compute the Maxwell stress averaged over time and angle $\langle w^{r\phi}\rangle$ as a function of $r$ from the full magnetic field in 3D in two ways:  by integrating $b^{\hat{r}}b^{\hat{\phi}}$ in Kerr-Schild coordinates and by integrating in a local frame comoving with the fluid with appropriate comoving volume element. (See~\cite{Krolik:2004ay,Noble:2010mm,Beckwith:2008pu} for details.)  We use the frame of the azimuthally averaged velocity for easier comparison with the stress from the azimuthal mean field.  [Of course, the azimuthally averaged velocity is itself a nonlocally-defined field, but once computed it has a local value at each point in 3D or 2D ($r$, $\theta$) space.]  Like in~\cite{EventHorizonTelescope:2019pcy}, we find that the two prescriptions give almost identical results.  As is commonly done, we define a Shakura-Sunyaev $\alpha_{\rm SS}$ by
\begin{equation}
  \alpha_{\rm SS} \equiv \frac{\langle w^{r\phi}\rangle}{\langle P_G+P_B\rangle}\;,
  \label{eq:alphass_from_avg}
\end{equation}
where $P_G$ and $P_B$ are the gas and magnetic pressure, respectively.
The result, included in Fig.~\ref{fig:alpha_ss}, closely resembles the left panel of Fig.~19 in~\cite{EventHorizonTelescope:2019pcy}, especially the dip at smaller $r$ in the lower resolution results, which is a new check on the consistency of IllinoisGRMHD with the codes with spherical-polar grids.  We also compute the Maxwell stress from the azimuthally averaged field $\mathbf{\overline{B}}$, using the same MHD formulas and in the azimuthally averaged comoving frame.  This is found to provide a much lower stress except very near the black hole, indicating that most of the angular momentum transport in the bulk of the disk accomplished by the magnetic field is done so by the non-axisymmetric field (the fluctuations, not the mean), a point also mentioned by Dhang~\etal~\cite{Dhang:2019kqo}.  Of course, using an ideal MHD formula for the stress tensor is not quite appropriate for the azimuthally averaged field tensor, which includes the effect of $\mathbf{\Delta E}$, but this residual EMF is itself a measure of the prevalence of nonaxisymmetry in the field and so does not affect the conclusion.

Because of the difference between the local 3D and azimuthally averaged velocity fields ($\mathbf{v}$ vs $\mathbf{\overline{v}}$), there is also a mean Reynolds stress which can transport angular momentum.  This is calculated in the frame comoving with the local azimuthally averaged velocity field:  $t^{\hat{r}\hat{\phi}} = \overline{\rho h u^{\hat{r}}u^{\hat{\phi}}}$.  (Note that the azimuthal mean velocity does not contribute, since if the field were axisymmetric, $u^{\hat{i}}$ would be zero in the frame comoving with the azimuthal average velocity.)  This Reynolds component is lower than the full (primarily non-axisymmetric) Maxwell component but still larger than the Maxwell stress of the mean field.

\begin{figure}
\includegraphics[width=1\columnwidth]{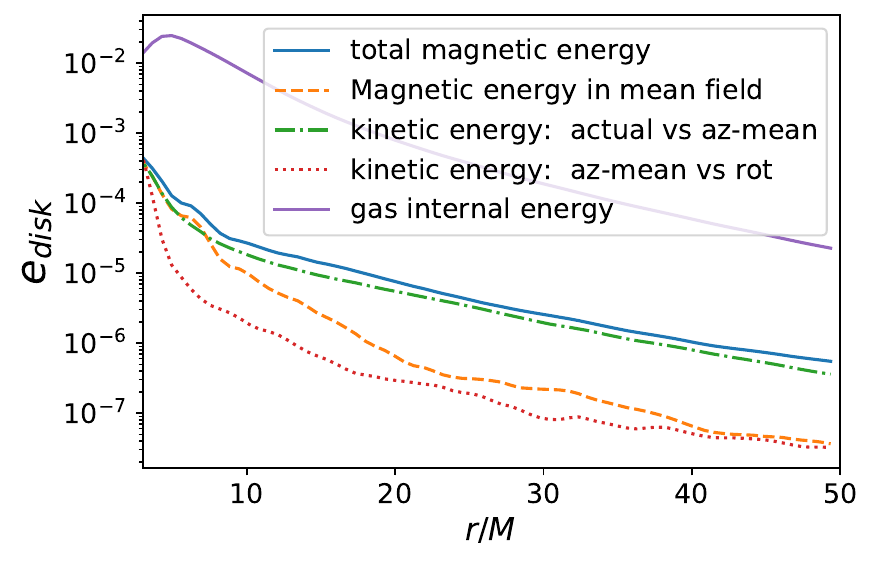}
\caption{Time and angle-averaged energy densities within the disk as functions of $r$.}
\label{fig:energy}
\end{figure}

In addition to the shear-type stresses quantified by $\alpha_{\rm SS}$, there can be a pressure-like contribution to the stress tensor from the non-axisymmeric Reynolds and Maxwell stresses.  We define the turbulent pressure as
\begin{equation}
P_t \equiv \frac{1}{3}\Delta T^{ab}(g_{ab}+u_au_b)\;.
\end{equation}
This is a very rough definition---applied instead to the stress-energy tensor of ideal MHD, it would give $P + b^2/6$, where $b^2=b_ab^a$ and $\mathbf{b}$ is the magnetic field in the $\mathbf{u}$ frame.  We find this pressure to be close to the magnetic pressure (of the full field), which even after saturation is about a factor of 10 lower than the gas pressure.  (See Fig.~\ref{fig:energy}, discussed below, which plots the associated energy densites.)  Thus, although turbulence dominates the angular momentum transport, it is very subdominant in pressure effects (e.g., determining the disk height).

We also compute time and angle-averaged energy densities in the disk.  For the full magnetic energy density, this is $b^2/2$, computed using the 3D local $\mathbf{B}$ and $\mathbf{u}$.  For the magnetic energy of mean field, we use the same formula (justified as above for stresses) using azimuthally averaged quantities.  The internal energy density of the gas is $P/(\Gamma-1)$.  The kinetic energy of a fluid with 4-velocity $\mathbf{u}_1$ relative to a frame $\mathbf{u}_2$ can be defined as follows.  Start from the Reynolds term in the fluid stress tensor $T_R^{ab}(\mathbf{u}_1)\equiv \rho h u_1^au_1^b$.  The desired energy is in the $\mathbf{u}_2$ direction, so we must project $T_R^{ab}u_{2a}$.  For an energy density, we must also specify a 4-velocity indicating the direction of flux of this energy we are considering; this is the frame in which the spatial integral converting energy density to energy would be performed.  The natural choices are $\mathbf{u}_1$ (for comoving frame) and $\mathbf{n}$ (for regular spatial integration on the slice), and we choose the latter.  The Reynolds energy density is then $u_{2a}n_{b}T_R^{ab}(\mathbf{u}_1)=-(\mathbf{u_1}\cdot \mathbf{u_2})W\rho h$.  We then subtract off the rest contribution when $\mathbf{u}_1=\mathbf{u}_2$ to get
\begin{equation}
e_K = -[(u_1\cdot u_2) + 1]W\rho h\;.
\end{equation}

There are three relevant 3-velocity fields in this problem: 1) $\mathbf{v}$, the 3D local velocity; 2) its azimuthal mean $\mathbf{\overline{v}}$; and 3) the rotational velocity, which is $\mathbf{\overline{v}}$ with radial and meridional components removed.  From each 3-velocity field, a 4-velocity field can be constructed by imposing the normalization $\mathbf{u}\cdot \mathbf{u}=-1$ (meaning that the $t$ component of the azimuthal mean 4-velocity is not exactly the azimuthal mean of the $t$ component of the local 4-velocity).  We thus compute a turbulent kinetic energy density, comparing the 3D local velocity to the azimuthal average, and a rotational kinetic energy, comparing the azimuthal average velocity to the rotational velocity.  Energies are plotted in Fig.~\ref{fig:energy}.  We omit the rotational energy, which dominates all others, as would be expected for an accretion disk.  The figure shows that the magnetic energy saturates more than an order of magnitude below the internal energy. However, the magnetic energy of the mean field saturates at a much lower value than this total magnetic energy.  This indicates that the magnetic energy is mostly non-axisymmetric (``turbulent''), and we infer that the energy in the turbulent magnetic field is close to the total magnetic energy.  The kinetic energy from the difference of azimuthal mean vs actual velocities (the turbulent kinetic energy) dominates over the non-rotational kinetic energy from the mean field, indicating that inside the disk eddy motion dominates over mean poloidal motions.  The total magnetic energy is seen to be very close to the turbulent kinetic energy, indicating that the kinetic and magnetic turbulent energies come into equipartition with each other.

\begin{figure}
\includegraphics[width=1\columnwidth]{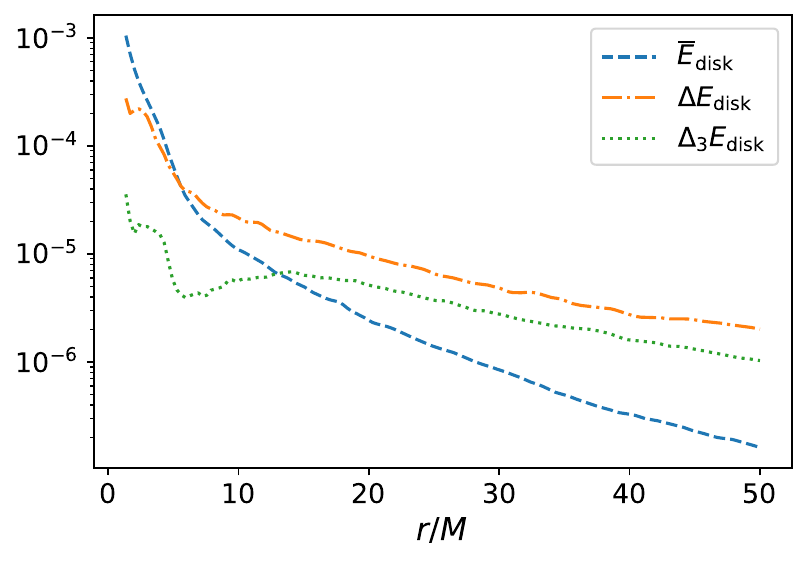} \\
\includegraphics[width=1\columnwidth]{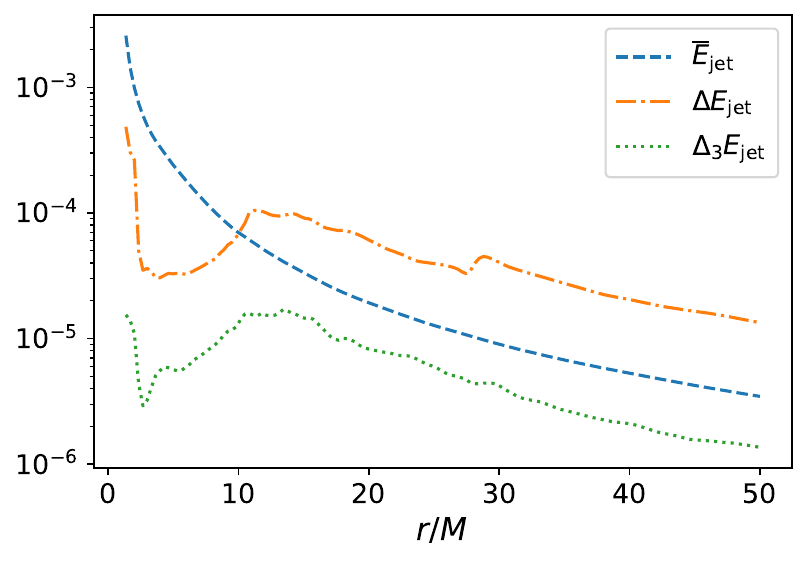}
\caption{Time and angle-averaged vector magnitudes of EMFs inside the disk as functions of $r$.  We compare the EMF of the mean fields with residual EMFs $\Delta E$ and $\Delta_3 E$.}
\end{figure}

\section{Isotropic dynamo models}
\label{sec:isotropic}

\subsection{Estimating an isotropic alpha effect}

If one knows that the dynamo EMF is of an isotropic alpha form, extracting the pseudoscalar coefficient \(\alpha_{\rm dyn}\) becomes straightforward.  Starting from the Newtonian formula $\mathbf{E}=\mathbf{B}\times\mathbf{v} + \alpha_{\rm dyn}\mathbf{B}$, one can take the dot product of both sides with respect to $\mathbf{B}$ to get $\alpha_{\rm dyn} = E\cdot B/B^2$.  Remembering that $B^2\gg E^2$ (see Fig.~\ref{fig:EBJ}), we can recognize this as the ratio of the two Lorentz invariants of the electromagnetic field.  We may, then, straightforwardly generalize the Newtonian formula as
\begin{equation}
\label{eq:alpha_dyn}
  \alpha_{\rm dyn} \equiv \frac{1}{2} \frac{{\star}F_{ab}F^{ab}}{F_{ab}F^{ab}}\;.
\end{equation}
a pseudoscalar, as expected.  Note that these are the invariants of the azimuthally averaged field tensors, not averages of the invariants of the field in 3D.

\begin{figure}
\includegraphics[width=1\columnwidth]{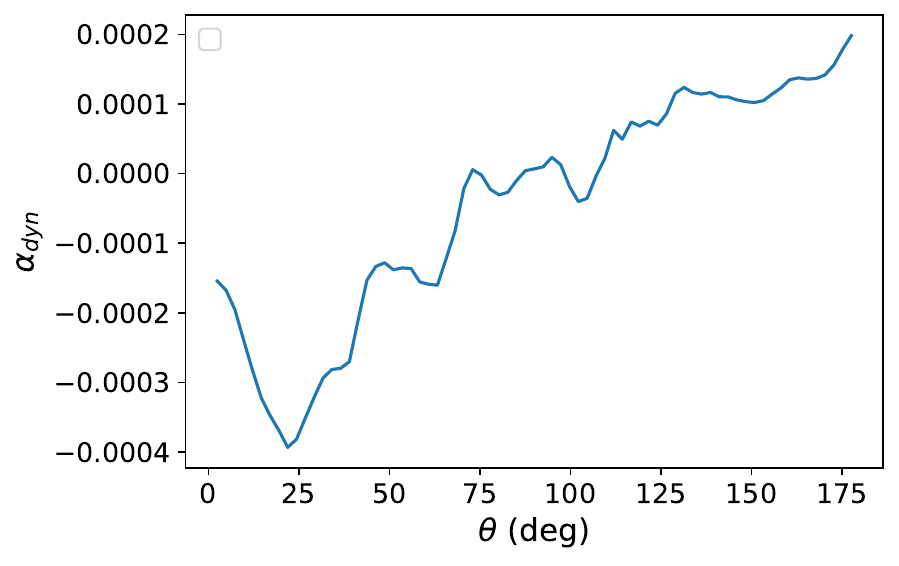}
\caption{Time and radial averaged $\alpha_{\rm dyn}$ (as defined by Eq.~\ref{eq:alpha_dyn} versus $\theta$.}
\label{fig:alpha_dyn}
\end{figure}

Because $\alpha_{\rm dyn}$ is expected to change sign across the equator, it would be inappropriate to compute an average over angles at a given radius, or an average over the entire grid (unless one were to presume this parity and perform a sign flip in one hemisphere); it is better to average over time and radius to extract an average $\alpha_{\rm dyn}$ as a function of $\theta$.  This is plotted in Fig.~\ref{fig:alpha_dyn}.

A hemispherical average pseudoscalar $\alpha_{\rm dyn}$ had previously been extracted from a 3D disk simulation by Hogg and Reynolds~\cite{Hogg:2018zon}.  They find $\alpha_{\rm dyn}$ in the range $1$--$2\times 10^{-4}$.  In magnitude, this is quite consistent with Fig.~\ref{fig:alpha_dyn}.  There has been some disagreement over the sign of $\alpha_{\rm dyn}$, with some finding it positive in the upper hemisphere and some finding it negative (always with the other hemisphere being opposite sign).  Our $\alpha_{\rm dyn}$ is negative in the Northern hemisphere, consistent with~\cite{Brandenburg:1995abc,davis2010sustained,Hogg:2018zon}.

\subsection{Calibrating isotropic diffusivity effect with 2D MHD}

To estimate an effective $\eta_{\rm dyn}$, we use the fact that the energy of $\overline{\mathbf{B}}$ inside the disk in the 3D simulation is smaller not only than the magnetic energy in the disk of the non-averaged field in 3D, but also than the magnetic energy for a 2D MHD simulation of the same disk.  One could say that the difference in energy is transferred from the mean field to the non-axisymmetric field.  Since this field is not captured by the fields tracked in a 2D simulation, it is in a sense equivalent to internal energy.  The nonaxisymmetric effect that accomplishes this effective dissipative transfer thus acts as a diffusivity, and one can estimate an effective $\eta_{\rm dyn}$ by finding what level of resistivity is needed to reduce the disk magnetic energy to the same degree.

We wish to implement a version of \Eqref{eq:relativistic_dynamo} with $\alpha^i_j=\alpha_{\rm dyn}\delta^i_j$ and $\eta^i_j=\eta_{\rm dyn} \delta^i_j$ in a relativistic 2D MHD code.  Such codes usually evolve the magnetic field 2-form (or, equivalently, the densitized magnetic field vector) ${}^2B_{ij} \equiv \epsilon_{ijk}B^k \equiv [ijk]\tilde{B}^k$.

Including an effective resistivity will introduce second spatial derivatives in the induction equation.  We wish to obtain a spatially covariant equation while avoiding Christoffel symbols to the extent possible.  Thus, we continue with form notation where possible, although the spatial metric necessarily will be invoked in Hodge duals and in creating the B-field 1-form ${}^1\mathbf{B}=B_a dx^a$.  We introduce an EMF 1-form $\mathbf{R}$ to include all deviations from ideal MHD.  Continuing in form notation, the dynamo-modified induction equation is
\begin{equation}
\partial_t\form{2}{\mathbf{B}}=-\mathcal{L}_{\mathbf{v}}\form{2}{\mathbf{B}} + d\mathbf{R}\;.
\end{equation}
For the standard isotropic dynamo, $\mathbf{R}$ is
\begin{equation}
  \mathbf{R}^{\rm dyn}=\alpha_{\rm dyn}\form{1}{\mathbf{B}} + \eta_{\rm dyn}\star d\,\form{1}{\mathbf{B}}\;,
\end{equation}
where in fact we use $\form{u}{\mathbf{B}}$ as defined in Eq.~\ref{eq:uBidealE} for the magnetic field.  In components,
\begin{eqnarray}
  R^{\rm dyn}_k &=& \alpha_{\rm dyn}B_k + \eta_{\rm dyn}\frac{1}{2}\epsilon_{ijk}A^{ij}\;, \\
  A^{ij} &=& \gamma^{i\ell}\gamma^{jm}(\nabla_\ell B_m - \nabla_m B_\ell) \nonumber \\
  &=& \gamma^{i\ell}\gamma^{jm}(\partial_\ell B_m - \partial_m B_\ell)\;.
\end{eqnarray}
(I.e., by using exterior derivatives, we avoid covariant derivatives.)  In order to avoid the timestep limitations of explicitly evolving a parabolic system, we promote $R_k$ to an evolution variable driven to its dynamo form:
\begin{equation}
  \partial_t R_k = \tau_{\rm drive}^{-1}(R^{\rm dyn}_k - R_k)\;.
\end{equation}
We set $\tau_{\rm drive}^{-1}=\hat{\tau}_d\Omega_K$, where $\Omega_K$ is the Keplerian angular frequency; we report results for $\hat{\tau}_d=0.5$ but have checked that the behavior of all quantities is insensitive to it within a wide range.  Smaller $\tau_{\rm drive}$ will force $R_k$ to track $R^{\rm dyn}_k$ more closely, but we wish to keep $\eta_{\rm dyn}/\tau_{\rm drive} < 1$ to avoid acausal propagation speeds.

The coefficient $\eta_{\rm dyn}$ is set to $c_s\ell$, where $\ell$ is an effective mixing length; it is natural to suppose that it is similar in magnitude to the mixing length used to set the alpha effective viscosity.  This similarity is confirmed for shearing box studies (more precisely, that the turbulent magnetic Prandtl number is roughly unity~\cite{Yousef:2003ey,Guan:2009dc,Fromang:2009cw}), but it should be remembered that the distinction between mean and turbulent field is different in these studies, so their applicability to azimuthally averaged dynamics cannot be certain.  We set
\begin{equation}
\label{eq:alpha_eta}
\ell = \alpha_{\eta} c_s\Omega_K^{-1}f_\rho f_\beta f_{\rm BH}\;,
\end{equation}
where $\alpha_{\eta}$ is a free constant (the symbol ``$\alpha$'' chosen for its analogy to $\alpha_{\rm SS}$, not to $\alpha_{\rm dyn}$), and the factors $f_\rho=\rho/(\rho+10^{-4}\rho_{\rm init,max})$, $f_\beta=P_{\rm gas}/(P_B+P_{\rm gas})$, $f_{\rm BH}={\rm max}[0, 1-(r/3M)^{-2}]$ suppress resistivity in the low-density region, the magnetosphere, and near the horizon, respectively.

We implement the above in the harmpi code, a version of the GRMHD code HARMPI~\cite{Gammie:2003rj,Noble:2005gf} publicly available at github~\cite{HARM2024}.  Note that, unlike truly 4D covariant resistive MHD codes, we do not evolve the actual electric field, although making $\mathbf{R}$ an evolution variable arguably accomplishes something similar.  Our less-precise implementation does have the advantage that no problem of stiffness arises in the limit of low $\eta_{\rm dyn}$.

\begin{figure}
\includegraphics[width=1\columnwidth]{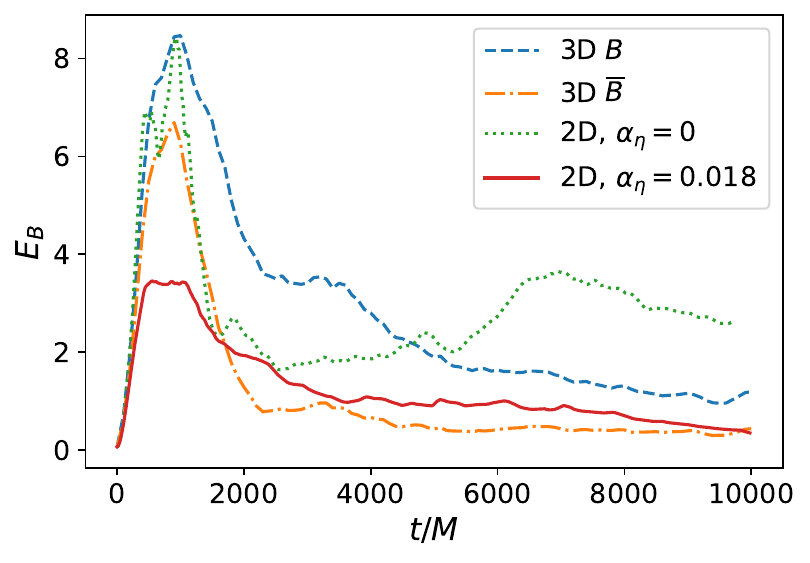} \\
\caption{Total magnetic energies in the disk vs time.  For the 3D run, we show total magnetic energy and magnetic energy in the mean field.  For 2D runs, we show runs with two different values of the effective resistivity parameter $\alpha_{\eta}$ (see \Eqrefp{eq:alpha_eta}).}
\label{fig:energy_vs_time}
\end{figure}

In Fig.~\ref{fig:energy_vs_time}, we show the magnetic energy in the disk as a function of time for several simulations.  For the 3D simulation, we plot both total magnetic energy and the (much smaller) magnetic energy of the mean field.  We also show harmpi in 2D with ideal MHD, which produces a magnetic energy far too high during the evolution period to match the mean of the 3D field.  Better agreement can be achieved by adding an isotropic dynamo $\eta$ term with $\alpha_{\eta}=0.018$.

\begin{figure}
\includegraphics[width=1\columnwidth]{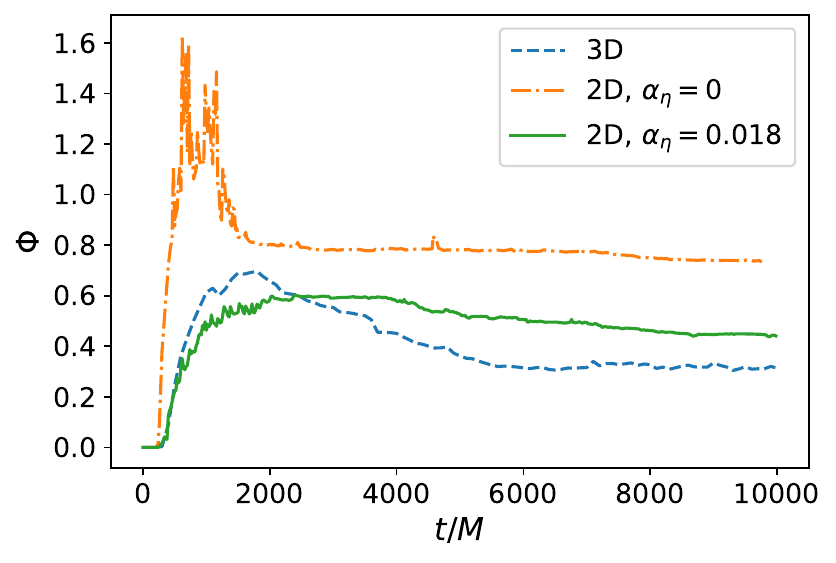} \\
\caption{Horizon fluxes vs time for the same runs as in Fig.~\ref{fig:energy_vs_time}.}
\label{fig:flux_vs_time}
\end{figure}

In Fig.~\ref{fig:flux_vs_time}, we plot magnetic flux $\Phi \equiv \int |B^r| dA$ on the horizon.  For electromagnetic energy extraction and jet formation, it is important that suppressing the disk magnetic energy does not also suppress the horizon flux.  Our implementation of $\eta$ does not add resistivity outside the high-density region or near the horizon, and we do see that the $\alpha_{\eta}=0.018$ run maintains a horizon flux that is fairly constant and similar to 2D.  It remains to be seen how it would fare for longer timescales (not pursued for this project for lack of a 3D comparison) and whether the polar field, which eventually diminishes in 2D, needs to be replenished by a slight alpha effect. Furthermore, it is a separate challenge to show that, for a more extended magnetized disk (e.g., the more common MAD test disks~\cite{Tchekhovskoy:2011zx}), magnetic flux could, over an extended period of time, accurately survive while advecting onto the horizon.

\section{Non-isotropic dynamo models}
\label{sec:nonisotropic}

We now attempt to extract $\alpha^i_j$ and $\eta^i_j$ by linear least-squares fitting.  Suppose one attempts a fit using $N$ data points, each data point corresponding to one event---a particular spatial location at a particular time.  At each data point, there are three equations, one for each component $\Delta E^i$.  For a given component $i$, combining all data points, we have the matrix equation that we wish to be satisfied
\begin{equation}
\mathbf{y}_i \cong \mathbf{A}_i\cdot\mathbf{x}_i\;, 
\end{equation}
where $\mathbf{y}_i$ is length-$N$ column vector made of all the $\Delta E^i$, $\mathbf{x}_i$ is a 6-dimension column vector made of the $\alpha^i_j$ and $\eta^i_j$ $\forall j$, and $\mathbf{A}_i$ is a $N\times 6$ matrix made of $B^j$ and $J^j$ $\forall j$.  We use the symbol ``$\cong$'' to mean that this is an equation deviation from which will be minimized (but, since there are more conditions than variables, will not be exactly satisfied).  The best fit model is a particular $\mathbf{x}_i=\mathbf{x}_i^{\rm fit}$ .

In choosing a residual function to be minimized, one must decide how data points are weighted.  One natural choice is to treat the absolute error at all points as equally important and minimize $(\mathbf{y}_i -\mathbf{A}_i\cdot\mathbf{x}_i^{\rm fit})^T
(\mathbf{y}_i -\mathbf{A}_i\cdot\mathbf{x}_i^{\rm fit})$.  This will have the effect that the fit will care more to reduce relative error at points with strong field and high $\Delta E^i$.  Alternatively, one might wish to weight relative errors at each point similarly.  One can do this by normalizing $\mathbf{A}$ and $\mathbf{y}$ at each spacetime event $P$ used in the fit, dividing all components of each by $||y||_P$, the L2 norm of the elements of $\mathbf{y}$:
\begin{equation}
||y||_P=\sqrt{\frac{1}{3}\sum_{i=1}^3y_i^2(P)}
\end{equation}
It is not obvious which sort of weighting is more desirable, so we have experimented with both.  For runs with weighting, we wish to avoid the possibility of fits dominated by attempts to reduce relative error in unimportant weakly-magnetized, very low-EMF regions.  Therefore, for these fits, we multiply $\mathbf{A}$ and $\mathbf{y}$ at each event $P$ by the weighting factor
\begin{equation}
\label{eq:weighting}
w_P = \max(||y||_P, 10^{-3} y_{\rm max})^{-1}\;,
\end{equation}
where $||y||_P$ is the L2 norm $||y||$ at $P$, and $y_{\rm max}$ is the maximum value of $||y||_P$ across all $P$ used for the given fit (i.e., the L$\infty$ norm across events of the L2 norm at each event).

\subsection{Model validation}

One measure of model accuracy is how well the model matches the original data.  We compute this accuracy by the following measure:
\begin{equation}
\chi_i^2 \equiv \frac{(\mathbf{y}_i -\mathbf{A}_i\cdot\mathbf{x}_i^{\rm fit})^T
(\mathbf{y}_i -\mathbf{A}_i\cdot\mathbf{x}_i^{\rm fit})}
{ \mathbf{y}^T\mathbf{y}}\;.
\end{equation}
This is related to the usual chi-squared statistic but with a different normalization.  More commonly, $\chi^2$ is normalized by the expected error, so that its magnitude for a good fit is unity.  In this case, we normalize by the the data itself, so a very good fit would have $\chi^2\ll 1$.  This is probably too much to hope for, so we would settle for $\chi^2<0.5$ or so.

We also measure the variance, which provides the uncertainty in the best fit values $\mathbf{x}_i^{\rm fit}$.  A variance estimate is provided by the SVD composition, valid for Gaussian statistics and uncorrelated data points.  In fact, even for the staggering in space and time we use to mitigate correlations, data points remain significantly correlated, so Bendre~\etal~\cite{bendre2020turbulent} find (and we confirm) that the Gaussian variance significantly underestimates the uncertainty in $\mathbf{x}_i^{\rm fit}$.

Instead, we estimate variance, similarly to Bendre~\etal, by computing $\mathbf{x}_i^{\rm fit}$ restricting the data used to different subsets of timesteps.  In our case, we compute $\mathbf{x}_i^{\rm fit}$ for even timesteps ($\mathbf{x}_{\rm even}$), for odd timesteps ($\mathbf{x}_{\rm odd}$), and for all timesteps ($\mathbf{x}_{\rm all}$).  Between two fits, $\mathbf{x}^A$ and $\mathbf{x}^B$, define relative difference for each component $i$ of $\mathbf{x}$ as $\delta^{A,B}_i\equiv (x^A_i - x^B_i)^2/(x^A_i + x^B_i)^2$.  For each $i$, we get an overall $\delta_i$ by taking the maximum of $\delta^{\rm even,\ odd}_i$, $\delta^{\rm even,\ all}_i$, and $\delta^{\rm odd,\ all}_i$.  For the fit to have some value, there must be at least one component of $\mathbf{x}^{\rm fit}$ that is consistent across fits (and this will probably be the one that contributes most to the fit), so for our overall variance measure, we minimize over components $i$:
\begin{equation}
{\rm var}(\mathbf{x}) = \min_{i}\bigl[\max(\delta^{\rm even,\ odd}_i, \delta^{\rm even,\ all}_i, \delta^{\rm odd,\ all}_i)\bigr]\;.
\end{equation}
Some fits involve more than one point in space, so for Models 2 and 4 below we also compute variances by skipping even and odd points in $r$ and (if it varies within the fit) $\theta$.  Then for each $x_i$ component we maximize over relative differences for different time and space samples before extracting the best-constrained component.

This is a very generous measure, since we only require one dynamo coefficient to be well constrained.  For the model parameters to be meaningful, we require ${\rm var}(\left[ \mathbf{x}\right])<0.3$.

Note that the more global the model, the more data points are relevant per fitting parameter.  This will tend to reduce variance, since one has very good statistics, but it is more difficult to have acceptable $\chi^2$, since the model must work well for more points.  Conversely, for more local models, $\chi^2$ will likely become smaller, but one should not be too encouraged by this, because one can fit anything as the number of conditions drops toward the number of fitting parameters, and the tendency to over-fit will often be revealed in a high variance.

\subsection{Pointwise models}

We attempt two models for which $\alpha^{i}_{j}$ and $\eta^{i}_{j}$ are taken to be functions of spatial coordinates but independent of time.  In {\bf Model 1}, they are taken to be functions of $r$ and $\theta$, i.e., there is an independent fitting procedure applied at each point.  This is similar to the model construction of Dhang~\etal~\cite{Dhang:2019kqo}.  In {\bf Model 2}, we take the dynamo coefficients to be functions of $\theta$, i.e., they are presumed to be constant in both $r$ and $t$.  Thus, compared to Model 1, each fit in Model 2 involves more data (corresponding to multiple points along a radial ray), which will tend to result in better variance but worse $\chi^2$.  We perform fits both with and without weighting, meaning equations at data points are weighted either with the factor 1 or the factor $w$ in \Eqref{eq:weighting}.

Results for model validation are shown in Table~\ref{tab:model12}.  For presenting average variance and $\chi^2$, we divide the space into two regions:  ``jet'' for the region within $\pi/3$ of the axis, and ``disk'' for all other latitudes.  As we see, none of the models are very good by either of our validation measures.

\begin{table}
    \begin{center}
    \caption[Models 1 and 2]{ Fit quality measures for models 1 and 2.}
    \label{tab:model12}
    \begin{tabular}{ccccc}\toprule
    Model & region & weighting & $\chi^2$  & var \\\midrule
    1  & disk & 1   &  0.599 & 0.789 \\
    & disk & $w$ & 0.776 & 0.834\\
    & jet & 1    & 0.483 & 0.532 \\
    & jet & $w$  & 0.747 & 0.781 \\\midrule
    2  & disk & 1   & 0.899 & 0.418 \\
    & disk & $w$ & 0.938 & 0.355 \\
    & jet & 1    & 0.985 & 0.206 \\
    & jet & $w$  & 0.969 & 0.328 \\\bottomrule
    \end{tabular}
    \end{center}
\end{table}

\subsection{Global models}
\label{sec:global_models}

We now attempt to fit to a model for which $\alpha^{ij}$ and $\eta^{ij}$ are not explicit functions of spatial coordinates.

For a covariant model, we would also have to build fitting formulas for these tensors out of physically relevant vectors and tensors, e.g., the metric $\gamma^{ij}$ or the velocity $v^i$.  Since $\alpha^{ij}$ is axial, we require at least one pseudoscalar or pseudovector ingredient.  The construction of an azimuthally averaged system only makes sense if an approximate azimuthal Killing vector $\mathbf{e_{\phi}}=\partial/\partial\phi$ has been identified.
We can, thus, assume its availability and expect that it may appear in the formulas for dynamo tensors, both because of its physical relevance (as roughly the direction of mean velocity) and its relevance in constructing the 2D system (which breaks 3D rotational invariance).  We will use the normalized vector
\begin{equation}
\mathbf{\hat{e}_{\phi}}\equiv (g_{\phi\phi})^{-1/2}\frac{\partial}{\partial\phi}\;.
\end{equation}
We get a pseudovector, corresponding to the direction of the symmetry axis, by taking a curl:
\begin{equation}
\zeta^i \equiv \epsilon^{ijk}\nabla_j(e_{\phi})_k = \epsilon^{ijk}\partial_j(e_{\phi})_k\;.
\end{equation}
We will use the normalized version $\mathbf{\hat{\zeta}}\equiv \mathbf{\zeta}/|\zeta|$, which resembles the Cartesian $z$ direction.

Some dynamo models also use $\nabla\rho$, the gradient of the density, or they use a gradient of the turbulent velocity (assuming some knowledge of this)~\cite{Brandenburg:1995abc}.
For a thin accretion disk, the direction of density gradient would be close to sign($z$)$\mathbf{\hat{\zeta}}$ except on the equator.  To build the expected parity factor in $\alpha_{\rm dyn}$ (something like $\cos\theta$), we instead want a nearly radial third vector, which could be interpreted as the direction of gravity.  Since we have already assumed an approximate timelike Killing vector is available to us, $\mathbf{e_t}$, we might as well use it to define this direction
\begin{eqnarray}
  \mathbf{f} &\equiv& -\nabla(e_t\cdot e_t) = -\nabla{g_{tt}}\;, \\
  \mathbf{\hat{f}}&\equiv& \mathbf{f}/|f|\;.
\end{eqnarray}
 This then provides the pseudoscalar $\mu \equiv \hat{f}\cdot\hat{\zeta}$, which is close to $\cos\theta$.  The resulting triad \{$\mathbf{\hat{\phi}}$, $\mathbf{\hat{\omega}}$, $\mathbf{\hat{f}}$\} provides a suitable vector basis inside the disk.  For global models including the polar region, it is not usable because on the poles $\mathbf{\hat{\omega}}$ and $\mathbf{\hat{f}}$ are parallel or anti-parallel.  We can alternatively use the azimuthal Killing vector to define
\begin{eqnarray}
  \mathbf{\varpi} &\equiv& -\nabla(e_{\phi}\cdot e_{\phi}) = -\nabla{g_{\phi\phi}}\;, \\
  \mathbf{\hat{\varpi}}&\equiv& \mathbf{\varpi}/|\varpi|\;.
\end{eqnarray}
Thus, $\mathbf{\hat{f}}$ resembles the spherical polar radial direction and is used to define the pseudoscalar $\mu$, while $\mathbf{\hat{\varpi}}$ resembles cylindrical polar radial direction and is used as a basis vector.

We then have a triad of normalized but not orthogonal vectors in $\mathbf{\hat{e}_{\phi}}$, $\mathbf{\hat{\zeta}}$, and $\mathbf{\hat{\varpi}}$ (and an associated dual cobasis); we denote components in this basis by hatted capital Latin letters, e.g. $B_{\hat{I}}=B\cdot e_{\hat{I}}$.

The global dynamo model is a tensor equation for constructing the residual EMF $\mathbf{\Delta E}$ in terms of the magnetic field $\mathbf{B}$, an axial vector, and some polar vector $\mathbf{Z}$, which we choose to be either $\mathbf{J}\equiv \mathbf{\nabla}\times \mathbf{B}$ or $B^2\mathbf{v}_T$, where $\mathbf{v}_T$ is the transport velocity of the fluid.  For $\mathbf{B}$, we report results using $\form{u}{\mathbf{B}}$ (and its curl for $\mathbf{J}$), although we have also tried using the normal frame $\mathbf{B}$ with similar results.

The tensor equation is
\begin{align}
  \Delta E^{\hat{I}} &= \alpha^{\hat{I}\hat{K}}B_{\hat{K}} + \eta^{\hat{I}\hat{K}}Z_{\hat{K}}\;, \\
  \alpha^{\hat{I}\hat{K}}(\mathbf{x},t) &= 
  a^{\hat{I}\hat{K}} \mu^{n_a(\hat{I},\hat{J}}) \psi(\mathbf{x},t)\;, \\
  \eta^{\hat{I}\hat{K}}(\mathbf{x},t) &= 
  e^{\hat{I}\hat{K}} \mu^{n_e(\hat{I},\hat{J})} \psi(\mathbf{x},t)\;,
\end{align}
where $n_a(\hat{I},\hat{J})$ and $n_e(\hat{I},\hat{J})$ are either 0 or 1, whichever is needed to obtain consistent parity for all terms.  

The scalar function $\psi(\mathbf{x},t)$ can be any scalar function constructed from the dynamical variables, with variables expected to be correlated with $\mathbf{\Delta E}$ being attractive choices.  Below, we consider several choices, including unity and the 2D spatial shear scalar $\sigma=(\sigma_{IJ}\sigma^{IJ})^{1/2}$.  We have also tried the gradient of the angular velocity $|\nabla\Omega|$, the sound speed $c_s$, and the fraction of the pressure that is gas or magnetic (i.e. $P_G/(P_G+P_B)$ and $P_B/(P_G+P_B)$).
We also try the scalar RMS deviation of $\mathbf{B}$ and of $\mathbf{v}_T$ from their azimuthal average, e.g.,
\begin{equation}
\label{eq:rms}
B_{\rm rms} = \left[\overline{|B-\overline{B}|^2}\right]^{1/2} \;.
\end{equation}
In an actual 2D simulation, these functions would not be available.  Rather, new evolution variables would have to be introduced with evolution equations designed to suitably approximate them.  However, designing such dynamics is only a worthwhile exercise if it is found that the actual RMS measures of non-axisymmetry provide useful $\psi(\mathbf{x},t)$ variables.

We also consider two alternatives for what constitutes the EMF residual:  $\Delta{\mathbf{E}}$ and $\Delta_m{\mathbf{E}}$.

We produce two types of global models. For {\bf Model 3}, we use data across all $r$, $\theta$, and $t$ to produce a single fit for the {\it numbers} $\alpha^{\hat{I}\hat{K}}$, $\eta^{\hat{I}\hat{K}}$.  For {\bf Model 4}, we use the time average of the data and fit for the numbers $\alpha^{\hat{I}\hat{K}}$, $\eta^{\hat{I}\hat{K}}$ using points across all $r$, $\theta$.

We introduce a couple of variations.  First, it is possible that the dynamics is very different in the jet and disk regions, so we create fits using data only within certain latitude regions.  For this purpose, ``disk'' is defined as the region between $\pi/3 < \theta < 2\pi/3$ and ``jet'' is defined as the regions $\theta<\pi/5$ and $\theta>4\pi/5$, so the intermediate coronal region is excluded from both.  This division is slightly different from the ``disk'' and ''jet'' designations for reporting Models 1 and 2 quality measures in Table~\ref{tab:model12}, but the separation here plays a different role.  It is used to determine which data is used for creating a given fit, rather than just which region is being averaged over for reporting a given norm after fitting is completed.

Also, one could argue that a dynamo model needn't capture small-scale features of $\Delta E^{\hat{I}}$, so in some cases we try smoothing data before fitting.  This effectively reduces the amount of independent data and can be expected to give higher $\chi^2$.  We convolve the data with a smoothing kernel, here chosen to be a triangle function in each direction centered on the point.  Specifically, if the half-length of the triangle is $N_{\rm sm}$ points, then variable $f$ at point $i$, $j$ (indexing location in $r$ and $\theta$, respectively) is smoothed as follows:
\begin{eqnarray}
  f_{i,j} &\rightarrow& \sum_{m=-N_{\rm sm}}^{N_{\rm sm}}\sum_{n=-N_{\rm sm}}^{N_{\rm sm}} \frac{w_{mn}}{\mathcal{Z}} f_{i+m,j+n}\;, \\
  w_{mn} &=& \left(1-\frac{|m|}{N_{\rm sm}+1}\right)\left(1-\frac{|n|}{N_{\rm sm}+1}\right)\;, \\
  \mathcal{Z} &=& \sum_{m=-N_{\rm sm}}^{N_{\rm sm}}\sum_{n=-N_{\rm sm}}^{N_{\rm sm}} w_{mn}\;.
\end{eqnarray}   
Thus, data at each point is replaced by a weighted sum of its neighbors, and the amount of smoothing is determined by the half-length of the triangle, $N_{\rm sm}$.

Quality of fit measures are reported in Table~\ref{tab:model34}. 
 Due to the large number of events used to fit a small number of global parameters, best-fit values to at least some dynamo coefficients can now be determined precisely. However, $\chi^2$ reveals that none of these best fits are good matches for the data, at least not good enough according to our criteria for viability.  We see many cases with $\chi^2$ close to 1, which occurs because the functions $\mathbf{B}$ and $\mathbf{Z}$ do not well overlap with $\mathbf{E}$ (see Fig.~\ref{fig:EBJ} and~\ref{fig:Evcor}), and none of our chosen correlates $\psi$ successfully corrects for this, so the best fit is obtained by making the model values $\mathbf{A}\cdot \mathbf{x}^{\rm fit}$ small compared to the data $\mathbf{y}$.

\begin{table}
    \begin{center}
    \caption[Models 3 and 4]{ Fit quality measures for models 3 and 4.}
    \label{tab:model34}
\begin{tabular}{cccccccS[table-format=1.3]S[table-format=1.5]}\toprule
    Model & region & $\mathbf{Z}$ & $w$ & $m_t$ & $\psi$ & $N_{\rm sm}$ & \multicolumn{1}{c}{$\chi^2$} & \multicolumn{1}{c}{var} \\\midrule
    3  & all & $J$   & 1   & 0  & 1             & 3  & 0.975  & 0.0068  \\
    & & $v$   & 1   & 0  & 1             & 3  &  0.970 & 0.018  \\
    & & $J$   & 1   & 0  & $v_{\rm rms}$ & 3  &  0.664 & 0.00124  \\
    & & $J$   & 1   & 0  & $\sigma$      & 3  & 1.000  &  0.001 \\
    & & $J$   & 1   & 3  & 1             & 3  &  0.996 & 0.00652  \\
    & & $J$   & 1   & 3  & $v_{\rm rms}$ & 3  & 0.980  & 0.00103  \\
    & & $v$   & 1   & 3  & $v_{\rm rms}$ & 3  &  0.978 & 0.00103  \\
    & & $J$   & $w$ & 0  & 1             & 3  &  0.992 & 0.057  \\
    & & $v$   & $w$ & 0  & 1             & 3  &  0.970 & 0.054  \\\midrule
    4 & jet & $J$   & 1   & 0  & 1             & 3  & 0.797  & 0.00096  \\
    & jet & $v$   & 1   & 0  & 1             & 3  &  0.549 & 0.0082  \\
    & disk & $J$   & 1  & 0  & 1             & 3  & 0.427  & 0.00285  \\
    & disk & $v$   & 1  & 0  & 1             & 3  &  0.473 & 0.0004  \\
    & all & $J$   & 1   & 0  & 1             & 3  &  0.854 & 0.0027  \\
    & & $v$   & 1   & 0  & 1             & 3  & 0.600  & 0.001  \\
    & & $J$   & 1   & 0  & 1             & 0  & 0.934  & 0.008  \\
    & & $J$   & 1   & 0  & 1             & 6  & 0.741  & 0.001  \\
& & $J$   & 1   & 0  & $\sigma$      & 3  &  0.934 & 0.0145  \\
    & & $J$   & 1   & 0  & $v_{\rm rms}$ & 3  & 0.685  & 0.001  \\
    & & $J$   & 1   & 0  & $B_{\rm rms}$ & 3  &  0.966 & 0.008  \\
    & & $J$   & $w$ & 0  & 1             & 3  & 0.922  &  0.001 \\
    & & $J$   & 1   & 3  & 1             & 3  & 0.934  & 0.001  \\
    & & $J$   & 1   & 3  & $B_{\rm rms}$ & 3  &  0.968 & 0.008 \\
    & & $v$   & 1   & 3  & 1             & 3  &  0.611 & 0.001 \\
    & & $J$   & 1   & 3  & $v_{\rm rms}$ & 3  &  0.771 & 0.001
    \\\bottomrule
\end{tabular}
    \end{center}
\end{table}

\section{Summary and conclusions}
\label{sec:conclusions}

Modeling non-axisymmetric effects in accretion disks is challenging because, for some quantities of interest, the non-axisymmetric/fluctuating part dominates over the  axisymmetric/mean part.  Inside the disk, the mean field is often itself more like residual noise than a dominant field component, and it is not surprising that non-axisymmetric effects cannot be easily correlated to its local values.  One crucial feedback of fluctuations onto the mean structure is the angular momentum transport, often captured via viscous hydrodynamics.  Mean fields remain important in the polar region, and for the long-term maintenance and evolution of this region, and for the late-time oscillating toroidal mean field, 2D models will need to model the mean field while incorporating non-axisymmetric effects.  Using results for a 3D Cartesian FMR GRMHD code, we have quantified the effective dynamo $\alpha_{\rm dyn}$ and $\eta_{\rm dyn}$.  We then tested whether this model is a good fit to the residual EMF, considering a wide range of pointwise and global models.  We have considered dynamo closures of growing complexity.  1) We considered models with $\mathbf{\Delta E}$ a function of $\mathbf{B}$ and $\mathbf{J}$, perhaps with added proportionality to some other function of variables available to the 2D mean system.  2) We generalized this to use proportionality to the RMS magnetic or kinetic energy, assuming that some evolution system involving only mean fields could be devised to roughly mimic these.  3) We generalized further and considered that only the high-$m$ residual $\mathbf{\Delta_3 E}$ need be fit by a closure involving mean fields.  None of these provided satisfactory models (although, of course, evolving $m=1$ to $m=3$ modes of all variables would capture a significant fraction of the non-axisymmetric effects in itself even without a closure for $\mathbf{\Delta_3 E}$, at the cost of being a low-azimuthal resolution 3D model rather than a 2D model).  In particular, designing evolution equations for $B_{\rm RMS}$ or $v_{\rm RMS}$ does not seem promising because we have failed to find a way to use them effectively.  More general and sophisticated models could still be attempted, e.g., providing evolution equations for 
$\mathbf{\Delta E}$ itself, analogous to turbulence models that evolve the Reynolds stress.

On the other hand, this is a stricter test than what is required of viscous hydrodynamics, in that it is usually not considered necessary to show that the residual stress tensor is well-fitted by a viscous stress tensor with some kinematic viscosity that is either a constant or is some specified function of azimuthal mean variables.  Rather, we insist that the viscosity capture one particular effect and calibrate a coefficient setting $\nu$ so that one effect has the appropriate magnitude.  Similarly, if one identifies the key features needed for long-term evolution, e.g., reducing poloidal field strength in the disk while maintaining it in the poles, simple models can be calibrated to do this. 

The main interest of this study is to have explored some new ways of quantifying dynamo effects in 3D simulations.  The main limitation of this study is the narrowness of the data to which it has been applied.  More could be learned by carrying out a similar analysis on a run that continues much longer for multiple dynamo cycles.  It is also necessary to run with different accretion disk-black hole configurations:  disks around lower-spinning black holes, different seed fields (e.g. toroidal), larger disks, disks that become magnetically arrested within a reasonable evolution time, radiatively efficient disks.

\begin{acknowledgments}

We acknowledge S.~Bose, M.~Forbes, F.~Foucart, F.~H.~Nouri, and E.~Most for useful conversations and encouragement, and we express deep gratitude to A.~Tchekhovskoy for making the HARMPI code publicly available.  M.D.\ gratefully acknowledges support from the NSF through grants PHY-2110287 and PHY-2407726 and support from NASA through grant 80NSSC22K0719. Z.B.E.\ gratefully acknowledges support from NSF awards AST-2227080, OAC-2227105, PHY-2110352, and PHY-2409654, as well as NASA awards ISFM-80NSSC21K1179 and TCAN-80NSSC24K0100. 
B.J.K.\ gratefully acknowledges support from NASA LISA Preparatory Science award 80NSSC24K0360; this material is also based upon work supported by NASA under award number 80GSFC21M0002.
Computational resources for performing the GRMHD simulations were provided by the WVU Research Computing Thorny Flat HPC cluster, which is funded in part by NSF OAC-1726534.
Post-simulation analysis was performed in part on the Pleiades cluster at the Ames Research Center, with support provided by the NASA High-End Computing (HEC) Program.

\end{acknowledgments}


\bibliography{apssamp}

\end{document}